\documentstyle[12pt,epsf]{article}
\begin{document}
\title{
{\bf Chiral Fermions and Multigrid}
}

\author{Artan Bori\c{c}i \\
        {\normalsize\it Paul Scherrer Institute}\\
        {\normalsize\it CH-5232 Villigen PSI}\\
        {\normalsize\it Artan.Borici@psi.ch}\\
}

\date{}
\maketitle

\begin{abstract}
Lattice regularization of chiral fermions is an
important development of the theory of elementary particles.
Nontheless, brute force computer simulations are very expensive,
if not prohibitive. In this letter I exploit the non-interacting
character of the lattice theory in the flavor space and propose
a multigrid approach for the simulation of the theory.
Already a two-grid algorithm saves an order of
magnitude of computer time for fermion propagator calculations.
\end{abstract}


\bigskip
\bigskip

\section{Introduction}
After many years of research in lattice QCD, it was possible
to formulate QCD with chiral fermions on the lattice
\cite{Kaplan,Shamir&Furman_Shamir,Narayanan_Neuberger,
Hasenfratz_Laliena_Niedermayer}.

The basic idea is an expanded flavor space which may be seen
as an extra dimension with 
left and right handed fermions defined in the two opposite
boundaries or walls.

Let $N$ be the size of the extra dimension, $D_W$ the Wilson-Dirac
operator, and $m$ the bare fermion mass. Then, the theory with
{\it Domain Wall} fermions is defined by the action
\cite{Kaplan,Shamir&Furman_Shamir}: 
\begin{equation}
\begin{array}{l}
S_{DW} := \bar{\Psi} {\cal{M}} \Psi = \sum_{i=1}^{N}
  \bar{\psi}_i [(D^{||}-1)\psi_i + P_{+}\psi_{i+1} + P_{-}\psi_{i-1}] \\
  P_{+}(\psi_{N+1} + m \psi_1) = 0, ~~P_{-}(\psi_0 + m \psi_N) = 0
\end{array}
\end{equation}
where $\cal{M}$ is the five-dimensional fermion matrix of the
regularized theory and
$D^{||} = M - D_W$ with $M \in (0,2)$ being a mass parameter.

I define also a theory with {\it Truncated Overlap Fermions} in complete
analogy with the domain wall fermions by substituting
\cite{Borici_LAT99}:
\begin{equation}
P_{+} \rightarrow (D^{||} + 1)P_{+}, ~~P_{-} \rightarrow (D^{||} + 1)P_{-}
\end{equation}

Both theories can be compactified in the walls of the extra
fifth dimension as low energy effective
theories given by the action:
\begin{equation}
\begin{array}{l}
S_{eff} = \bar{\chi}_1 D \chi_1 \\
\chi_1 = P_{+}\psi_1 + P_{-}\psi_N,
 ~~\bar{\chi}_1 = \bar{\psi}_1 P_{-} + \bar{\psi}_N P_{+}
\end{array}
\end{equation}
where $D$ is the chiral Dirac operator satisfying the
Ginsparg-Wilson relation \cite{Ginsparg_Wilson}:
\begin{equation}\label{GWR}
\gamma_5 D^{-1} + D^{-1} \gamma_5 = a \gamma_5 R,
\end{equation}
where $a$ is the lattice spacing and $R$ is a local operator
trivial in the Dirac space.

I defined Truncated Overlap Fermions such that in the large $N$
limit one obtains Overlap Fermions \cite{Narayanan_Neuberger}
with the Dirac operator given by \cite{Neuberger1}:
\begin{equation}
D_{OV} = \frac{1 + m}{2} - \frac{1 - m}{2} \gamma_5 \mbox{sgn}(H)
\end{equation}
where $H = \gamma_5 D^{||}$.

Untill now, the computations with chiral fermions with
the standard algorithms have been very
expensive. The extra fermion flavors introduce a large overhead.
One multiplication with the fermion matrix costs ${\cal{O}}(n)$
$D_W$-multiplications with $n \sim N$ for domain wall fermions and
much larger for the overlap operator
\cite{Borici,Neuberger2,Edwards_Heller_Narayanan}.

In this letter I propose a multigrid algorithm along the
fifth dimension which makes these simulations much faster.
The key observation is the lack of gauge connections along
this dimension. It is well-known that the overhead
of such algorithms scales like $N$log$N$.

Here it is the $multigrid$ algorithm:
{\it ALGORITHM1 (Generic) for solving the system $D_{OV} x = b$:}
\begin{equation}
\begin{array}{l}
\mbox{Given} ~N, ~x_0, ~r_0(=b), ~tol, ~tol_1, ~\mbox{set} ~tol_0 = 1
~\mbox{and iterate:} \\
for ~i = 1, \ldots \\
~~~~tol_0 = tol_0 tol_1 \\
~~~~\mbox{Solve} ~D y = r_{i-1} ~\mbox{within} ~tol_0 \\
~~~~x_i = x_{i-1} + y \\
~~~~r_i = b - D_{OV} x_i \\
~~~~if ||r_i||_2 < tol, ~end ~for \\
\end{array}
\end{equation}
where by $o$ is denoted a vector with zero entries and $tol_1,tol$
are tolerances. $tol_1$ is typically orders
of magnitude larger than $tol$ such that the work per $D_{OV}$
inversion is minimized.

{\it Remark 1}. The straightforward application of the $ALGORITHM1$
gives a two-grid algorithm. By calling it
again in solving the smaller system and iterating,
one gets a full multigrid algorithm.

{\it Remark 2}. The corresponding Hybrid Monte Carlo (HMC) algorithm can be
obtained by working with an approximate Hamiltonian in the coarse lattice
and by a global correction on the fine lattice.

In Fig. 1 we compare the norm of the residual $r_i = b - D_{OV}x_i$
of the Conjugate Residual (CR) algorithm (which is optimal since $D_{OV}$
is normal \cite{Borici_thesis}) and $ALGORITHM1$. I gain about an order
of magnitude (in average) on $30$ $4^4$-configurations at $\beta = 6.0$
and $m = 0.1$.

For the coarse lattice we used $N = 6$ with the Truncated Overlap
Fermions and the Lanczos method to compute $D_{OV}$ \cite{Borici}.

Let $P$ be the following
orthogonal transformation:
\begin{equation}
\chi_i = P_{+} \psi_i + P_{-} \psi_{i-1}
\end{equation}
I computed the inverse of $D$ in the $ALGORITHM1$ using
\begin{equation}
D^{-1} = (P^{T} {\cal{M}}^{-1} {\cal{M}}_1 P)_{1,1}
\end{equation}
where ${\cal{M}}_1$ is the same as $\cal{M}$ but with $m = 1$
\cite{Borici_LAT99} and the subscript $1,1$ stands for the
$(1,1)$ block of an $N$x$N$ partitioned matrix along the fifth
dimension.

Recently, the possibility of a Multigrid algorithm along the
all dimensions is raised \cite{Rebbi_LAT99}. In this case a
gauge fixing is needed.

I would like to thank Philippe de Forcrand for suggestions
on how to improve the $ALGORITHM1$
and Herbert Neuberger for interesting discussions after my talk at
Lattice99 conference.

The author thanks PSI where this work was done and SCSC Manno
for the allocation of computer time on the NEC SX4.

\pagebreak

\begin{figure}
\epsfxsize=12cm
\epsfxsize=10cm
\vspace{3cm}
\centerline{\epsffile[100 200 500 450]{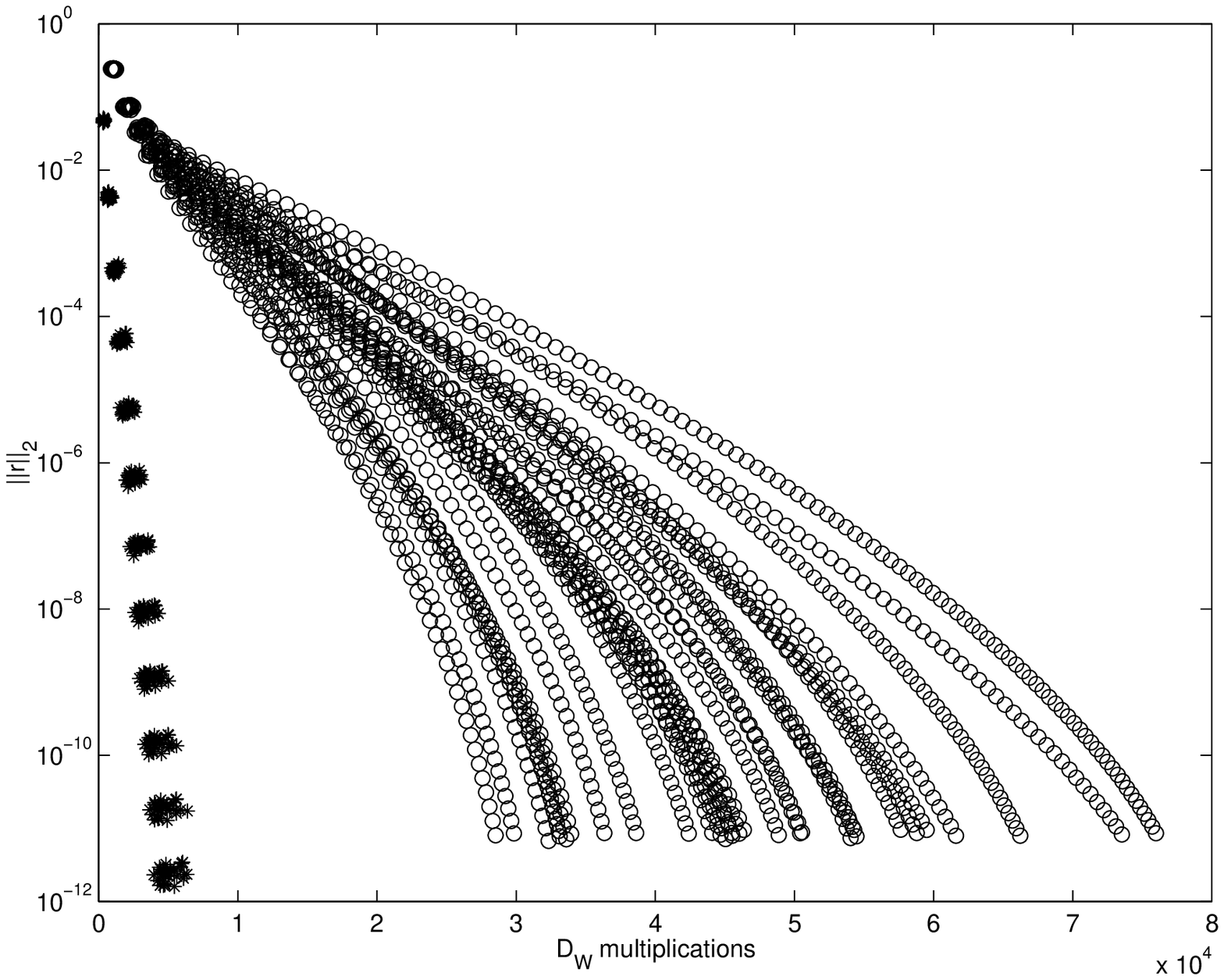}}
\caption{Norm of the residual error vs. the number of $D_W$
multiplications on 30 configurations. Circles stand
for the straightforward inversion with CR and stars for the $ALGORITHM1$.}
\end{figure}

\end{document}